\documentclass[revsymb,bibnotes,aip,apl,amsmath,amssymb,reprint,superscriptaddress,floatfix]{revtex4-1}

\usepackage[]{graphicx}
\usepackage{bm}
\usepackage{float}
\usepackage{lineno}
\usepackage{hyperref}
\usepackage[all]{hypcap}
\usepackage{chngcntr}

\usepackage{textgreek}
\usepackage{lipsum}
\usepackage{slashed}
\usepackage{amsmath}
\usepackage{amsfonts}
\usepackage{amssymb}%
\usepackage{balance}
\usepackage{color}
\usepackage{textcomp}

\makeatletter
\newcommand{\colorcaption}[2][]{%
  \begingroup%
  \renewcommand{\@caption@fignum@sep}{ (Color online). }%
  \caption[#1]{#2}%
  \endgroup%
}
\makeatother

\begin{document}

\author{M. Bejarano}
\affiliation{Helmholtz-Zentrum Dresden-Rossendorf, Institute of Ion Beam Physics and Materials Research, Bautzner Landstrasse 400, 01328 Dresden, Germany}
\author{F. J. T. Goncalves}\email{f.goncalves@hzdr.de}
\affiliation{Helmholtz-Zentrum Dresden-Rossendorf, Institute of Ion Beam Physics and Materials Research, Bautzner Landstrasse 400, 01328 Dresden, Germany}
\author{M. Hollenbach}
\affiliation{Helmholtz-Zentrum Dresden-Rossendorf, Institute of Ion Beam Physics and Materials Research, Bautzner Landstrasse 400, 01328 Dresden, Germany}
\affiliation{Technische Universit\"{a}t Dresden, 01062 Dresden, Germany}
\author{T. Hache}
\affiliation{Helmholtz-Zentrum Dresden-Rossendorf, Institute of Ion Beam Physics and Materials Research, Bautzner Landstrasse 400, 01328 Dresden, Germany}
\affiliation{Institut f\"{u}r Physik, Technische Universit\"{a}t Chemnitz, 09107 Chemnitz, Germany}
\author{T. Hula}
\affiliation{Helmholtz-Zentrum Dresden-Rossendorf, Institute of Ion Beam Physics and Materials Research, Bautzner Landstrasse 400, 01328 Dresden, Germany}
\affiliation{Institut f\"{u}r Physik, Technische Universit\"{a}t Chemnitz, 09107 Chemnitz, Germany}
\author{Y. Berenc{\'e}n}
\affiliation{Helmholtz-Zentrum Dresden-Rossendorf, Institute of Ion Beam Physics and Materials Research, Bautzner Landstrasse 400, 01328 Dresden, Germany}
\author{J. Fassbender}
\affiliation{Helmholtz-Zentrum Dresden-Rossendorf, Institute of Ion Beam Physics and Materials Research, Bautzner Landstrasse 400, 01328 Dresden, Germany}
\author{M. Helm}
\affiliation{Helmholtz-Zentrum Dresden-Rossendorf, Institute of Ion Beam Physics and Materials Research, Bautzner Landstrasse 400, 01328 Dresden, Germany}
\affiliation{Technische Universit\"{a}t Dresden, 01062 Dresden, Germany}
\author{G. V. Astakhov}
\affiliation{Helmholtz-Zentrum Dresden-Rossendorf, Institute of Ion Beam Physics and Materials Research, Bautzner Landstrasse 400, 01328 Dresden, Germany}
\author{H. Schultheiss}
\affiliation{Helmholtz-Zentrum Dresden-Rossendorf, Institute of Ion Beam Physics and Materials Research, Bautzner Landstrasse 400, 01328 Dresden, Germany}

\title{Mapping the stray fields of a nanomagnet using spin qubits in SiC}

\begin{abstract}
We report the use of optically addressable spin qubits in SiC to probe the magnetic stray fields generated by a ferromagnetic microstructure lithographically patterned on the surface of a SiC crystal. The stray fields cause shifts in the resonance frequency of the spin centers. The spin resonance is driven by a micrometer-sized microwave antenna patterned adjacent to the magnetic element. The patterning of the antenna is done to ensure that the driving microwave fields are delivered locally and more efficiently compared to conventional, millimeter-sized circuits. A clear difference in the resonance frequency of the spin centers in SiC is observed at various distances to the magnetic element, for two different magnetic states. Our results offer a wafer-scale platform to develop hybrid magnon-quantum applications by deploying local microwave fields and the stray field landscape at the micrometer lengthscale.
\end{abstract}
\maketitle

Optically active vacancy-related qubits, among others, the nitrogen-vacancy (NV) center in diamond and the silicon-vacancy ($\mathrm{V_{Si}}$) center in silicon carbide (SiC), are considered as nano-scale magnetic field sensors~\cite{Degen:2008jh, Kraus:2013vf} due to their long spin coherence time even at room temperature~\cite{Balasubramanian:2009fu, Simin:2017iw}. A hybrid quantum platform consisting of such spin qubits and nanostructured magnetic circuits can bring new functionalities and analytics tools~\cite{Balasubramanian:2008cz, Maze:2008cs, Simin:2016cp}. By using state-of-the-art lithography techniques, magnetic nanostructures can be tailored such that the resulting magnetic dipolar field landscape locally affects the spin qubits. While isolated nanostructures act locally, they can also be built into periodic patterns such that the magnetic dipolar fields are repeated over a certain distance\cite{Tacchi2012,Davies2015,Goncalves2016,Li2017,Iacocca2020}. Importantly, patterning of nanomagnets also allows control over the dynamic magnetic excitations, or spin waves, opening the possibility to coherently control spin qubits at the low power consumption~\cite{Andrich:2017ey} limit, characteristic to spintronic and magnonic devices~\cite{Hoffmann2014}. Spin waves can propagate over distances up to several millimeters\cite{Conca2013,Dubs2017,Flacke2019}, can be steered\cite{Davies2015a,Vogt2014,Wagner2016,Heussner2017}, focused\cite{Hoffmann2014} and can also modify the polarization of electromagnetic waves. One application in particular would be the local amplification of microwave fields acting on the SiC spin centers using spin waves~\cite{Kikuchi2017, Andrich:2017ey, Lee-Wong2020}. From a different standpoint, spin qubits can be used to sense magnetic excitations with high spatial resolution~\cite{Maertz:2010ko, Steinert:2010kk, Pham:2011dc, Rondin:2013kk}. For instance, this would allow the search for radiating microwave fields in spin orbit torque based nano-oscillators~\cite{Divinskiy2018,Hache2020}.

Though SiC is a natural material of choice for the aforementioned hybrid quantum platform at the wafer scale~\cite{Castelletto2020,Son:2020kh}, most of the efforts are put so far on the NV centers in diamond~\cite{Degen:2008jh, Balasubramanian:2009fu, Balasubramanian:2008cz, Maze:2008cs, Andrich:2017ey, Maertz:2010ko, Steinert:2010kk, Pham:2011dc, Rondin:2013kk}. \par In this letter, we demonstrate the use of spin centers created by proton irradiation in SiC as sensors for the local fields surrounding a magnetic nanostructure. The proton irradiation was done such that a high density of spin color centers was bound to the surface closest to the magnetic element to ensure optimal sensitivity. The spin resonance of these centers was driven by a patterned micrometer-sized microwave antenna such that the microwave signal is delivered locally and confined to the region of interest. Compared to conventional millimeter-sized microwave circuits, the patterned antennas were found to be more efficient at driving the resonance of the spin centers due to the scaling of the circuit dimensions. We find that the spatial profile of the magnetic dipolar fields produced by the magnetic element varies with the applied field direction and the magnetic state. 

\begin{figure}[t]
\centering
\includegraphics[width=8.5cm]{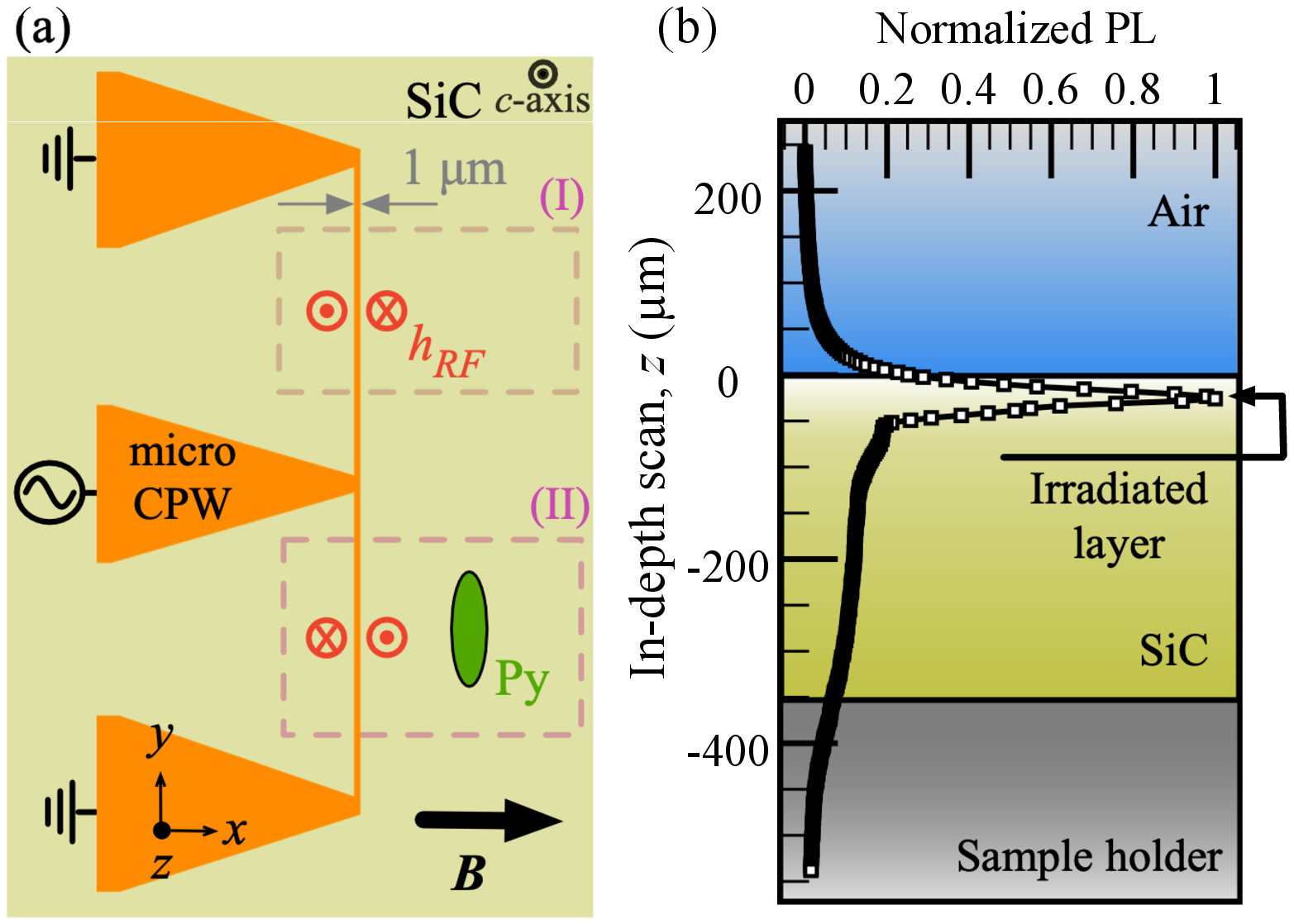}
\colorcaption{(a) Schematic illustration of the microwave circuit and the magnetic element fabricated on the SiC substrate. The symbols labelled as $h_{RF}$ illustrate a time snapshot of the out-of-plane microwave fields. Both the micro-CPW and the magnetic element were patterned on the immediate vicinity of the proton-irradiated SiC layer. The magnetic element is a 8~\textmu m $\mathrm{\times}$ 2~\textmu m ellipse. (b) In-depth scan of the PL (normalised) across the SiC substrate measured in region (I).}
\label{fig:FIG1}
\end{figure}

We perform proton irradiation on a high purity semi-insulating 4H-SiC substrate with a fluence of $10^{15}$~cm$^{-2}$ ~\cite{Kraus:2017cka}. The energy of the irradiated protons was 30~keV and corresponds to a mean projected range of 200~nm below the surface. The optically active spin centers are excited with a 785~nm laser and the detected photoluminescence (PL) in the spectral range from $850 \, \mathrm{nm}$ to $1050 \, \mathrm{nm}$ is characteristic for the $\mathrm{V_{Si}}$ centers in SiC\cite{Riedel:2012jq}. After proton irradiation, the SiC substrate went through two electron beam lithography (EBL) steps that enabled the fabrication of the micrometer-sized microwave circuit and the magnetic element. The microwave circuit consists of sputtered Cr~(6~nm)/Au~(150~nm) layers and the geometry corresponds to that of a shorted co-planar waveguide antenna (micro-CPW) as illustrated in Fig.~\ref{fig:FIG1}(a). The magnetic element is a 8~\textmu m $\mathrm{\times}$ 2~\textmu m ellipse consisting of 50~nm thick $\mathrm{Ni_{81}Fe_{19}}$ (Py). In the schematic of Fig.~\ref{fig:FIG1}(a) we highlight two regions of the micro-CPW: region (I), where there is no influence from the magnetic element ;and region (II), where the ellipse was patterned such that the effect of the magnetic dipolar fields could be probed. A separation of 40~\textmu m between the two regions ensured that region (I) did not experience any magnetic dipolar fields originating from the magnetic element. It is important to note that the efficiency and direction of the microwave excitation field is the same at both regions (I) and (II). Considering the lateral position of the microwave circuit with regard to the probed regions and the close proximity in height to the proton-irradiated SiC layer, one expects the microwave fields to be predominantly in the out-of-plane direction. \par 

Figure~\ref{fig:FIG1}(b) shows an in-depth PL scan across the substrate, measured in region~(I). Here, one can see that the proton-irradiated layer yields a considerably larger PL. On all experiments carried out, the $z$-position of the objective focal plane was chosen such that PL acquisition was maximised. The PL was acquired using a Si avalanche photodetector (Thorlabs APD440A) connected to a lock-in amplifier whose modulation frequency was set via amplitude modulation of the probing laser. In order to enhance the spatial resolution, a spatial filter with a 30~\textmu m pinhole was added to the detection path such that the measured PL originated only from a fraction of the volume illuminated by the incident laser. Thus, allowing spatial scans with a lateral resolution of 1.30~\textmu m, 1.65~\textmu m and 28~\textmu m in the $x$, $y$ and $z$ directions, respectively.

Optically detected magnetic resonance (ODMR)~\cite{Kraus:2013vf} was employed to measure the intrinsic resonance properties of the spin centers in region (I) as well as the effect of the magnetic element in region (II). This method is based on the relative change of the PL intensity, $\mathrm{\Delta PL / PL}$, under spin resonance conditions~\cite{Kraus:2013vf} (i.e. driven by external microwave signals). In this particular measurement setup, the ODMR spectra were obtained via the same lock-in detection method as the PL, except that the microwave signal was modulated instead of the laser. The main focus of our experiments was the limit of strong magnetic fields $\gamma B \gg 2D $ since it is known that even at zero external magnetic field the element will yield relatively strong magnetic dipolar fields. Here, $\gamma$ is the gyromagnetic ratio for the electrons captured by the $\mathrm{V_{Si}}$ centers and $2D$ is a zero field splitting in the $\mathrm{V_{Si}}$ centers. In 4H-SiC, the later can take the values $4.5 \, \mathrm{MHz}$ or $70 \, \mathrm{MHz}$, depending on the crystallographic configuration \cite{Kraus:2013vf, Nagy:2019fw}. For $|\gamma|\mathrm{=28~MHz / mT}$ which is the same as for an isolated electron, the above condition of the strong magnetic field is fulfilled for $B \gg 2.5\, \mathrm{mT}$. In this limit the spin resonance frequencies shift linearly with $B$ following $\nu_{1,2} = \gamma B \pm D (3 \cos^2 \theta -1)$ \cite{Kraus:2013di, Simin:2015dn}. The splitting between two spin resonances $\nu_{1} - \nu_{2}$ depends on the angle $\theta$ between the magnetic field and the symmetry axis of the spin center. For the case of pristine samples, the symmetry axis is parallel to the $c$-axis of the 4H-SiC lattice. In the context of the present experiments, the static magnetic field is applied perpendicular to the c-axis and the driving microwave field is applied along the c-axis.\par

\begin{figure}[t]
\centering
\includegraphics[width=8.5cm]{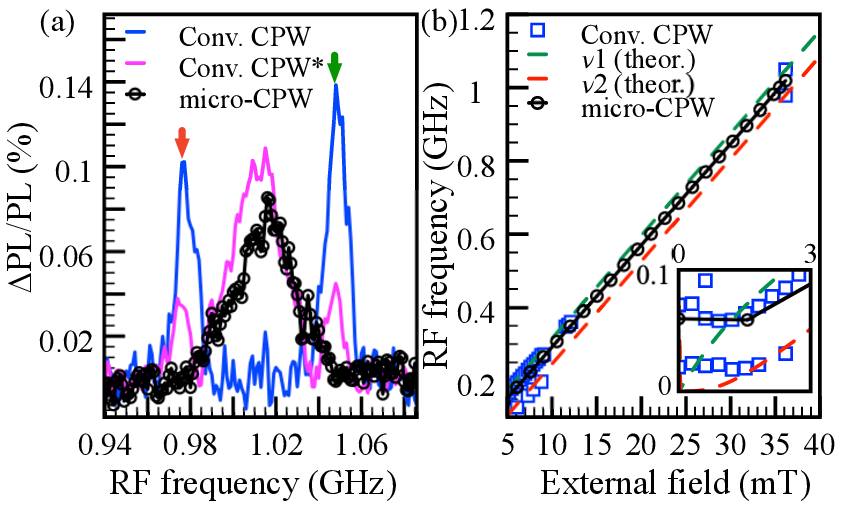}
\colorcaption{(a) ODMR spectrum of the proton-irradiated layer in SiC (black) obtained using the micro-CPW and of the pristine SiC bulk substrate (blue) measured using the conventional CPW as the source of microwave driving fields. The line in magenta shows the ODMR spectrum obtained while using the conventional CPW with the substrate on a flip-chip configuration. The external magnetic field was set to 36~mT and the data acquired at room temperature. (b) Field dependence of the ODMR frequency peak obtained following a Lorentzian fit to the experimental data. The averaged error in fitting the resonances was 0.5~MHz. The dashed lines $\nu_{1}$ and $\nu_{2}$ show the spin resonance frequencies which are solutions to the effective Hamiltonian of spin-3/2 centers in SiC for in-plane magnetic fields in the pristine SiC crystal\cite{Kraus:2013di}. The inset plot shows the near-zero field region where the ODMR peaks were observed at 60~MHz (micro-CPW), 20~MHz and 71~MHz (conventional CPW).}
\label{fig:FIG2}
\end{figure}
Figure~\ref{fig:FIG2}(a) shows the ODMR spectra of the proton-irradiated SiC layer obtained using the micro-CPW as the source for the driving microwave fields, at a constant in-plane field of 36~mT. The output power of the microwave source was set to 4~dBm and the electrical contact with the micro-CPW was done use a ground-signal-ground (GSG) probe. The resulting spectrum, shown in black, exhibits one resonance peak at 1.018~GHz with a linewidth of $\mathrm{37.0~\pm~0.5~MHz}$. As a control measurement, we acquired the ODMR spectrum originating from the bulk of the SiC substrate using a commercially available CPW with a signal line width of 275~\textmu m as the source of the driving microwave fields. On this particular experiment, the output power of the microwave source had to be increased to 25~dBm in order to obtain an ODMR amplitude comparable to that obtained using the micro-CPW. The corresponding spectrum is also shown in Fig.~\ref{fig:FIG2}(a) (blue) and exhibits two resonance peaks at 0.978~GHz and 1.048~GHz, with linewidths of 13~MHz. The spectra acquired using these two different excitation methods should have yielded similar results but clearly differ in both the number of peaks and the frequency linewidth. In order to understand the difference between these two spectra we present a third ODMR spectrum which was also obtained using the conventional CPW as the source of the microwave driving fields, except this time the substrate was positioned in a flip-chip configuration. This configuration allowed to achieve the shortest possible distance between the proton-irradiated SiC layer and the CPW, ultimately resulting in stronger driving microwave fields. This ODMR spectrum (magenta line in Fig.~\ref{fig:FIG2}(a)) revealed three resonance peaks, the two satellite peaks appear to coincide with those originating from the bulk of the SiC substrate, while the broader centre peak coincides with that of the proton-irradiated SiC layer. This result suggests that the difference in lineshape is caused by the proton irradiation fluence chosen for this particular SiC substrate.

Intense proton irradiation leads to inhomogeneous broadening due to local changes in the crystalline environment around each $\mathrm{V_{Si}}$ center~\cite{Kasper:2020dw}. Both, the $2D$ zero-field splitting and the orientation of the symmetry axis can be affected. As a consequence, a broad resonance peak is observed instead of the double peak structure characteristic of the pristine SiC crystal. Nevertheless, the single ODMR resonance observed shifts linearly with magnetic field applied perpendicular to the $c$-axis as shown by the field dependence plotted in Fig.~\ref{fig:FIG2}(b). From the fit of the experimental data we obtained a slope of $\mathrm{27.4 \pm 0.2~MHz/mT}$. This value is close to the gyromagnetic ratio of an isolated electron and also very similar to that observed for the two resonance peaks originating from the bulk SiC crystal. A more detailed discussion on the effect of the irradiation fluence on the ODMR spectra of SiC will be discussed elsewhere. Having identified the characteristic ODMR spectra of the proton-irradiated SiC layer, we proceed to demonstrate the spatial mapping of the magnetic dipolar fields produced by the magnetic element in region (II) by assessing the frequency shifts on the ODMR spectra as a function of the $x$ and $y$-positions relative to the magnetic element.

\begin{figure}[t]
\centering
\includegraphics[width=8.5cm]{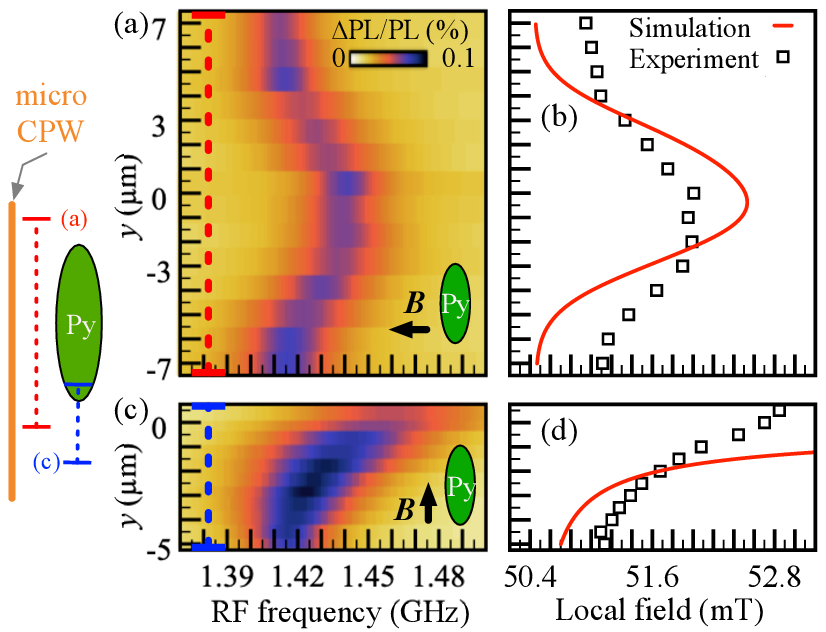}
\colorcaption{(a) ODMR spectra acquired at various positions parallel to the long axis of the ellipse. The scan indicated by the red dashed line was located 4~\textmu m away from the edge of the micro-CPW and 2~\textmu m away from the edge of the ellipse. A magnetic field of 50.5~mT was applied in-plane, along the short axis of the ellipse, as illustrated in the inset schematic on the left. (b) Plot of the magnetic dipolar fields obtained the micromagnetic simulations (line) and the magnetic field obtained from the experiments (markers). The local field in the vicinity of the ellipse corresponds to the external field added to the magnetic dipolar field contribution, amounting to 52.0~mT near $y$= 0 and 51.0~mT in the outer positions ($\mathrm{\pm}$~7~\textmu m) where the magnetic dipolar field contribution is very small. The error associated with the conversion to units of magnetic field in (b) and (d) is 0.3~mT. (c) ODMR spectra acquired at various positions collinear with the long axis of the ellipse (see inset schematic on the left). The scan line (blue dashed line) was located 7~\textmu m away from the edge of the micro-CPW. The in-plane magnetic field (50.5~mT) was applied parallel to the long axis of the ellipse. (d) Comparison between the local fields measured in the experiments and the magnetic dipolar fields obtained from the micromagnetic simulations.}
\label{fig:FIG3}
\end{figure}

Figure~\ref{fig:FIG3}(a) shows the ODMR spectra measured at various positions along the longest axis of the ellipse\footnote{For easier visualisation, the raw data represented in the color plots of Figs.3(a),(c) and 4(a) correspond to fitted Lorentzian functions (each frequency spectrum was smoothed and fitted using a Lorentzian lineshape). The fitting errors were very small compared to the measured quantities, justifying our approach. The average fitting errors in were: 3(a) frequency peak 0.5 MHz, peak amplitude $\mathrm{10^{-3}}$, FWHM 2~MHz.; 3(c) frequency peak 0.3~MHz, peak amplitude $\mathrm{10^{-5}}$, FWHM 0.3~MHz; 4(a) frequency peak 0.4 MHz, peak amplitude $\mathrm{10^{-3}}$, FWHM 0.7~MHz.}. Here, an in-plane magnetic field of 50.5~mT is applied perpendicular to the long axis of the magnetic element. Consider the schematic on the left hand side of Fig.~\ref{fig:FIG3}(a) as an illustration of the scanning path (dashed red line). The scanning line was parallel to the micro-CPW in order to ensure a constant microwave excitation efficiency. The microwave power was set to 0~dBm in this experiment. At the outer positions of the ellipse ($\mathrm{\pm}$ 7~\textmu m) the ODMR frequency is 1.412~GHz and increases towards 1.442~GHz as the shortest distance to the ellipse is reached ($y$=~0). This frequency shift is a direct consequence of the field gradient caused by the magnetic dipole distribution within the magnetic element. Taking into consideration the fitting parameters obtained from the data shown in Fig.~\ref{fig:FIG2}(b), one can convert the frequency shifts to local fields as plotted in Fig.~\ref{fig:FIG3}(b). From here we estimate a local field of 50.9~mT at the outer edges of the line scan and a field value of 52.0~mT at the centre of the magnet. On this same plot, we included the magnetic dipolar field distribution obtained via micromagnetic simulations (using Mumax3~\cite{arne2014}), for this particular sample geometry\footnote{Simulation size: 20~\textmu m by 14~\textmu m by 50~nm and magnetic parameters used: $M_s$=~810~kA/m, $A_{ex}$=13~pJ/m, damping $\alpha$=~0.007, $\gamma$= 29.6 GHz/T }. At 50.5~mT, where the magnetic ground state of the ellipse is a nearly saturated state, the spatial dependence of the generated magnetic dipolar fields are in good agreement with that obtained from the experiments.\par 
Let us consider a scenario where the external magnetic field is applied parallel to the long axis of the ellipse. This case is discussed in Fig.~\ref{fig:FIG3}(c), where the ODMR spectra are plotted as a function of the distance to the pole of the ellipse (see illustration on the left). In this experiment the microwave power was increased to 13~dBm to compensate for the weaker driving fields due to a larger distance to the micro-CPW. The ODMR frequency appears to follow a near asymptotic behaviour as the ellipse is approached, which highlights the possibility to achieve very strong local magnetic fields near the edges of the magnetic element. The increase in the resonance linewidths from 30~MHz ($y$= -5~\textmu m) to 67~MHz ($y$= -0.5~\textmu m) can be explained by the larger magnetic dipolar field gradients when probing closer to edge of the magnetic element. The reduction in the amplitude near the edge of the ellipse ($y$= 0) is due to the partial overlap between the probing laser beam and the magnetic material. Moreover, we anticipate that the lateral resolution ($\mathrm{\sim1.65}$~\textmu m) imposed by the dimensions of the PL acquisition volume also affects the agreement with the magnetic dipolar field profiles shown in Figs.~\ref{fig:FIG3}(b) and (d). Therefore, further developments can be made in terms of spatial resolution in order to increase the overall field sensitivity beyond the sub-millitesla limit which is currently achieved~\cite{Degen:2008jh,Balasubramanian:2008cz,Maze:2008cs}.\par
\begin{figure}[t]
\centering
\includegraphics[width=8.5cm]{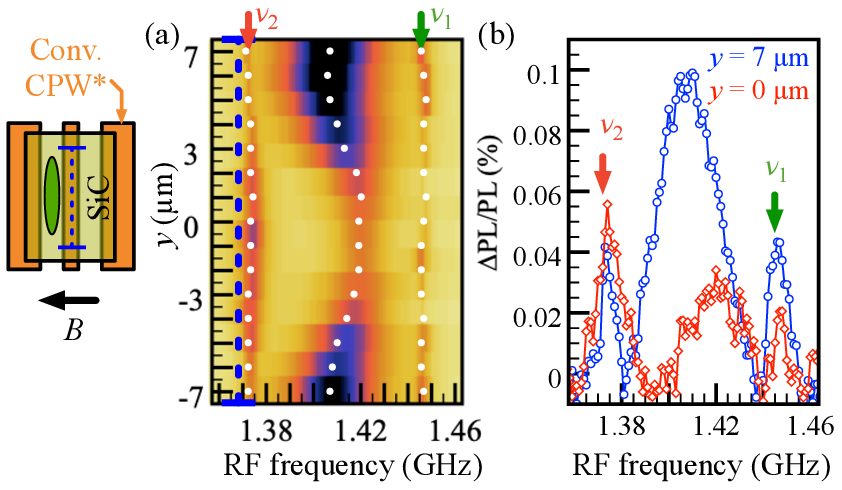}
\colorcaption{(a) ODMR spectra acquired at various positions along the longest axis of the ellipse, as illustrated by the blue dashed line shown in the schematic on the left. White markers correspond to fits to the spectra using three Lorentzian lineshapes. Note that only the broad centre peak shifts as a function of scanning position. (b) Examples of the ODMR lineshape obtained at 7~\textmu m and 0~\textmu m. The narrow satellite peaks have approximately the same frequency as those expected for $\nu_{1}$ and $\nu_{2}$, shown in Fig.~\ref{fig:FIG2}(b).}
\label{fig:FIG4}
\end{figure}

 
In order to further demonstrate the degree of localisation achieved using the lithographically defined micro-CPW and magnetic elements in combination with the proton-irradiated layer, we performed a control experiment where we mapped the magnetic dipolar fields of the ellipse using the conventional CPW, in a flip-chip configuration. The layout of this experiment is shown in Fig.~\ref{fig:FIG4}(a). We note that the magnetic element was placed in the gap between the ground and the signal line of the CPW which means that the driving microwave fields are predominantly oriented out-of-plane, as in the case of the micro-CPW. The spatial mapping of the ODMR spectra is shown in Fig.~\ref{fig:FIG4}(a). Similarly to the data shown in the inset of Fig.~\ref{fig:FIG1}(a), we observe two side resonance peaks and a broader center resonance peak. Note that only the center peak is shifting in frequency when varying the $y$-position. This means that while the SiC spin centers are being uniformly excited across the depth by the driving microwave fields, only the irradiated layer in close proximity to the ellipse is sensing the local magnetic dipolar fields.
Figure~\ref{fig:FIG4}(b) shows the ODMR spectra acquired at $y$= 7~\textmu m~(blue) and 0~\textmu m (red) positions of the scan line. The difference in ODMR peak position between these two modes yielded the largest frequency shift observed in this experiment (14~MHz). This frequency shift differs from that observed using the micro-CPW (30~MHz, see Fig.~\ref{fig:FIG3}) due to loss of spatial resolution as the spatial filter was removed in this experiment. In addition, it is worth noting that in order to detect the ODMR spectra shown here, the microwave input power had to be increased to 25~dBm, which is larger than that used in the micro-CPW measurements, by a factor of 315.\par 
In conclusion, we have demonstrated the use of SiC spin centers as sensors for the magnetic dipolar fields generated by a lithographically defined magnetic nanostructure. The two magnetic states we investigated generate distinct local magnetic dipolar field gradients, demonstrating that both the shape of the magnetic element and the applied field direction can be used to locally modify the resonance of the spin centers in SiC. The integration of the microwave circuit contributed to an increase in sensitivity compared to the use of millimeter-sized CPWs due to a more efficient delivery of microwave signal driving the spin resonance.\par The concepts we explore and the results obtained here may be directed, for example, towards the manipulation of the spin centers in SiC via the dynamic magnetic dipolar fields generated by spin waves in magnetic nanostructures.
\newline
This work was supported by the German Research Foundation (DFG) under the grants SCHU 2922/4-1 and AS 310/5-1. The lithography was done at the Nanofabrication Facilities (NanoFaRo) at the Institute of Ion Beam Physics and Materials Research, HZDR. Support from the Ion Beam Center (IBC) at Helmholtz-Zentrum Dresden-Rossendorf (HZDR) is gratefully acknowledged for the proton irradiation.
\newline
The data that support the findings of this study are available from the corresponding author upon reasonable request.
%

\end{document}